# Quasi-One-Dimensional Electronic Nature of Ta$_4$SiTe$_4$ Underlying the Giant Thermoelectric Performance

*Hironari Isshiki[1,2]\*, Ayumu Tabata[3], Masafumi Horio[2], Motoi Kimata[1], Youichi Yamakawa[4], Ryutaro Okuma[2], Miki Imai[2], Fumiya Matsunaga[4], Koshi Takenaka[4], Kenichi Ozawa[5], YoshiChika Otani[2], Fumio Komori[2], Iwao Matsuda[2], Masahiro Hara[6], Yoshihiko Okamoto[2], and Masayuki Hashisaka[2]*

[1] Advanced Science Research Center, Japan Atomic Energy Agency, 2-4 Shirakata, Tokai-mura, Naka-gun, Ibaraki, Japan

[2] Institute for Solid State Physics, University of Tokyo, 5-1-5 Kashiwanoha, Kashiwa, Chiba, Japan

[3] Graduate School of Science and Technology, Kumamoto University, 2-39-1 Kurokami, Chuo-ku, Kumamoto, Japan

[4] Department of Applied Physics, Nagoya University, Furo-cho, Chikusa-ku, Nagoya, Japan

[5] Institute of Materials Structure Science, High Energy Accelerator Research Organization (KEK), 1-1 Oho, Tsukuba, Ibaraki, Japan

[6] Faculty of Advanced Science and Technology, Kumamoto University, 2-39-1 Kurokami, Chuo-ku, Kumamoto, Japan

\*Corresponding Author: isshiki.hironari@jaea.go.jp

**ABSTRACT**: $Ta_4SiTe_4$ is a one-dimensional van der Waals material that exhibits an exceptionally large thermoelectric power factor below room temperature. However, since this material has been available only in the form of acicular microcrystals, experimental exploration of the electronic properties responsible for its giant thermoelectric performance has long been challenging. In this study, we quantitatively evaluated the one-dimensional electronic nature of $Ta_4SiTe_4$ by combining micro-spot angle-resolved photoemission spectroscopy and transport measurements on focused-ion-beam-processed samples. The angle-resolved photoemission spectroscopy measurements reveal anisotropic band dispersions along and perpendicular to the crystallographic *c* axis. Consistently, transport measurements demonstrate that the resistivity perpendicular to the *c* axis is approximately five times larger than that along the *c* axis at 200 K. These results provide direct experimental evidence for the quasi-one-dimensional electronic character of $Ta_4SiTe_4$, which underlies its giant thermoelectric response reported previously, and offer fundamental insights into the role of electronic dimensionality in enhancing thermoelectric performance.

**MAIN TEXT**: Achieving high thermoelectric performance requires simultaneously maximizing the Seebeck coefficient and electrical conductivity. However, these two quantities are generally constrained by a trade-off, as captured by the Mott relation in conventional band transport, limiting the development of high-performance thermoelectric materials[1,2]. One effective strategy to overcome this limitation is to exploit the electronic anisotropy inherent in low-dimensional materials, thereby enhancing thermoelectric performance along specific crystallographic directions[3–5]. Indeed, a variety of low-dimensional systems, such as $SnSe$[6], $Bi_2Te_3$[7,8], $CsBi_4Te_6$[9,10], and $Ta_2PdSe_6$[11–13], have demonstrated excellent thermoelectric performance. Despite these advances, the reduced dimensionality and characteristic crystal morphology of such materials often impose geometrical constraints on experiments, making it difficult to directly probe their electronic states and to examine a quantitative correlation with thermoelectric properties. Consequently, experimental approaches capable of accessing intrinsic electronic states independent of sample geometry are crucial for advancing the design principles of high-performance thermoelectric materials.

Among such low-dimensional thermoelectric materials, $Ta_4SiTe_4$[14,15], a one-dimensional van der Waals compound, has attracted significant attention due to a remarkably large thermoelectric power factor below room temperature, where only a limited number of materials exhibit high thermoelectric performance[16]. Figure 1(a) illustrates the crystal structure of $Ta_4SiTe_4$, which consists of one-dimensional chain-like units. Each chain consists of a periodic array of Si atoms, where each Si atom is located at the center of an antiprismatic cage formed by eight Ta atoms and further surrounded by eight Te atoms. In the bulk crystal, these chains are periodically arranged into a triangular lattice through van der Waals interactions. A previous study on bulk $Ta_4SiTe_4$ sample reported measurements along the $c$ axis, revealing a large negative Seebeck coefficient of $S \approx -400$ µV·K$^{-1}$ and a low electrical resistivity of $\rho \approx 2$ mΩ·cm at 100–200 K[16]. These values correspond to a large power factor $PF = S^2 \cdot \rho^{-1} \approx 80$ µW·cm$^{-1}$·K$^{-2}$, which significantly exceeds that of the representative thermoelectric material $Bi_2Te_3$ at room temperature (~30 µW·cm$^{-1}$·K$^{-2}$)[7,8]. Notably, partial substitution of Ta with Mo further enhances the power factor to approximately 170 µW·cm$^{-1}$·K$^{-2}$ in the temperature range of 220–280 K[16]. In addition, Ti substitution at the Ta site converts the original n-type conduction of $Ta_4SiTe_4$ to p-type, suggesting its potential for low-temperature thermoelectric applications[17]. High thermoelectric performance has also been reported for composites of $Ta_4SiTe_4$ with organic conductors, highlighting its promise among flexible thermoelectric materials[18,19]. Overall, $Ta_4SiTe_4$ constitutes an attractive platform for further investigations, including its theoretically predicted one-dimensional topological nature[20].

This outstanding performance of Ta$_4$SiTe$_4$ has been attributed to a quasi-one-dimensional electronic structure, as has been discussed for other low-dimensional materials. However, experimental verification of the band dispersion, transport anisotropy, and narrow band gap remains lacking, mainly because Ta$_4$SiTe$_4$ is available only in the form of whisker crystals with diameters on the order of a few hundred micrometers (see Fig. 1(b) and (c)). For example, electronic anisotropy in Ta$_4$SiTe$_4$ has so far been inferred only from optical reflectivity measurements[21] and has not been directly confirmed by electrical transport experiments. In this study, motivated by the need for direct and quantitative electronic characterization of low-dimensional thermoelectric materials, we combine micro-spot angle-resolved photoemission spectroscopy (micro-ARPES) using a small-diameter beam with transport measurements on focused-ion-beam (FIB)-processed micro devices to elucidate the electronic origin of the giant thermoelectric performance of Ta$_4$SiTe$_4$.

The whisker crystals of Ta$_4$SiTe$_4$ were synthesized by vapor-phase crystal growth[16,21]. Stoichiometric amounts of Ta and TeCl$_4$ powders, together with 100% excess Si powder, were mixed and sealed in an evacuated quartz tube. The tube was heated to 873 K for 24 h, then to 1423 K for 96 h, and finally furnace-cooled to room temperature. The addition of excess Si powder prevents spontaneous ignition when opening the quartz tube. The synthesized crystals were stored in a vacuum desiccator. In this study, we used Ta$_4$SiTe$_4$ crystals with diameters larger than 100 μm, as shown in Figs. 1(b) and (c).

The micro-ARPES was performed with a focused synchrotron radiation with the beam spot size of about φ10 μm at the beamline BL-28A of Photon Factory (KEK)[22]. To obtain a clean crystal surface, a rod was attached perpendicular to the $c$ axis of a whisker crystal (~100 μm in diameter), and *in situ* cleavage of the crystal was achieved by breaking the rod under ultrahigh vacuum (< 2 × $10^{-8}$ Pa). ARPES measurements were performed at a temperature of 20 K with a photon energy of 90 eV and total energy resolution of 30 meV. The sample was oriented such that the mirror plane defined by the incident light and the photoelectron detection direction was nearly parallel to the crystallographic $c$ axis. Subsequent X-ray diffraction measurements confirmed that the cleaved surface corresponds to the (100) plane.

Figure 2(a) presents the band dispersion of the $Ta_4SiTe_4$ along the $\bar{\bar{\Gamma}}$–$\bar{\bar{Z}}$ direction obtained by ARPES, where $\bar{\bar{\Gamma}}$ and $\bar{\bar{Z}}$ denote the high-symmetry points Γ and Z projected onto the $k_z$ direction (parallel to the $c$ axis), as illustrated in the right panel of Fig. 2(a). The clearly observed band dispersions are in good agreement with the results of previous first-principles band calculations including spin-orbit coupling, which are overlaid as black lines in the figure[16,20]. A constant-energy map in the $k_y$–$k_z$ plane at $E = E_F − 0.2$ eV, where $E_F$ is the Fermi energy, is shown in Fig. 2(b). The steep band dispersion along the crystallographic $c$ axis gives rise to high carrier mobility and low electrical resistivity, while the flat dispersion perpendicular to the $c$ axis results in a sharp energy dependence of the density of states near the Fermi level, leading to a large Seebeck coefficient. The anisotropic electronic structure of $Ta_4SiTe_4$ is therefore expected to overcome the conventional trade-off between electrical conductivity and the Seebeck coefficient, yielding an exceptionally large power factor. Figure 2(c) shows the band structure near the Fermi level along the direction indicated by the red line in Fig. 2(b). Figure 2(d) plots the photoemission intensity as a function of energy, from which the valence-band maximum is determined to be located at $E =$ $E_F − 0.18$ eV. Since the conduction band is not accessible in the present ARPES measurements, the magnitude of the band gap cannot be directly determined here, which will be discussed later based on the electrical transport measurements.

To date, the electrical resistivity of bulk $Ta_4SiTe_4$ has been measured only along the crystallographic $c$ axis[21]. In the present study, we fabricated multiterminal devices from needle-like crystals using an FIB technique, which allows us to measure electronic transport both along and perpendicular to the $c$ axis and thus to investigate the one-dimensional transport anisotropy. During device fabrication, a $Ta_4SiTe_4$ crystal was milled using a $Ga^+$ ion beam with a beam current of 50 nA and an acceleration voltage of 30 kV. A block with dimensions of 40 μm × 40 μm × 4.4 μm, with one edge aligned parallel to the $c$ axis, was extracted from a clean crystal surface, as indicated by the dashed outline in Fig. 1(c). The block was transferred onto a sapphire substrate and electrically connected to pre-patterned Au electrodes by Pt deposition in the FIB system. The block was then further milled to form a multiterminal device with final dimensions of $l = 15.0$ μm, $w = 4.0$ μm, and $t = 4.4$ μm as shown in Fig. 3(a). Throughout the fabrication process, the $Ga^+$ ion irradiation time was minimized to suppress unintended carrier doping that could influence the transport measurements.

Electronic transport measurements were carried out using a standard lock-in technique at a frequency of 93 Hz with the terminal configuration T1–T9 shown in Fig. 3(a). Direct current measurements were also performed and yielded results identical to those obtained by the lock-in method (not shown). For the longitudinal resistivity measurements, we injected an excitation current of $I = 10$ μA root-mean-square (rms) from terminal T1 toward the electrically grounded terminal T8. The resistivity along the $c$ axis was evaluated from the voltage measured between T2 and T3, while that perpendicular to the $c$ axis was obtained from the voltage between T4 and T5.

For the Hall resistivity, on the other hand, we injected $I$ = 100 μA rms and measured the Hall voltage $V_H$ between T7 and T9 while sweeping an out-of-plane magnetic field $H_z$. Measurements were performed on three independently fabricated devices, all of which yielded essentially identical results. Here, we show representative data obtained from one device.

Figure 3(b) shows the temperature dependence of the measured electrical resistivity parallel ($\rho_{//}$, black) and perpendicular ($\rho_\perp$, red) to the $c$ axis. For comparison, the two-terminal resistivity of a bulk $Ta_4SiTe_4$ crystal measured along the $c$ axis is also plotted ($\rho_{bulk}$, blue). The overall behavior of $\rho_{//}$ is consistent with $\rho_{bulk}$, including the broad maximum near 200 K, except at the lowest temperatures where surface electron doping induced by the FIB process becomes significant. Quantitatively, $\rho_{//}$ is slightly larger than $\rho_{bulk}$ over a wide temperature range, which may reflect uncertainties in the geometric factor in the previous bulk measurement. At room temperature, the resistivity anisotropy $\rho_\perp/\rho_{//} \approx 7$ is comparable to the anisotropy ≈ 8 inferred from low-energy optical conductivity measurement[21]. In the temperature range around 200 K, where $Ta_4SiTe_4$ exhibits high thermoelectric performance, the resistivity anisotropy decreases to $\rho_\perp/\rho_{//} \approx$ 5.3.

Figure 4(a) shows the Hall resistivity $\rho_H = (V_H/I) \cdot t$ as a function of magnetic field $H_z$ at various temperatures. Here, $t$ = 4.4 μm is the thickness of the sample. The negative slope of the measured $\rho_H$ indicates that the dominant charge carriers are electrons, consistent with the negative Seebeck coefficient reported for this material[23]. The electron doping is likely attributed to tellurium vacancies[17]. Assuming that the electrical transport is dominated by electrons, the n-type carrier density $n$ at each temperature was evaluated by a linear fit to the Hall data using $n = \mu_0 H_z \cdot (e\rho_H)^{-1}$, where $\mu_0$ and $e$ are the vacuum permeability and the elementary charge, respectively. The temperature dependence of the n-type carrier density is shown in Fig. 4(b), exhibiting a semiconductor-like temperature dependence.

Here, we estimate the band gap $E_g$ from the thermally excited carrier density. In a semiconductor, the intrinsic carrier density $n_i$ is given by

$$n_i \propto T^{\frac{3}{2}} \cdot \exp\left(-\frac{E_g}{2k_B T}\right), \quad (1)$$

where $k_B$ is the Boltzmann constant. Accordingly, Eq. (1) can be rewritten as

$$\ln\left(\frac{n_i}{T^{3/2}}\right) = -\frac{E_g}{2k_B}\left(\frac{1}{T}\right) + \text{const.} \quad (2)$$

Figure 4(c) shows $\ln\left(\frac{n}{T^{3/2}}\right)$ plotted as a function of 1/$T$. According to the DFT calculations, the curvature of the conduction band in this material is larger than that of the valence band[16], indicating that the effective mass of electrons is smaller than that of holes. Consequently, the Hall response is expected to be dominated by electrons even in the thermally activated regime. Based on this consideration, we estimate the band gap from the Hall measurement results by neglecting the contribution from holes. Assuming that the temperature range $T \geq 240$ K corresponds to an intrinsic-like (thermally activated) regime, in which the plot becomes approximately linear, the data are well fitted by Eq. (2) (green line). From this fit, the band gap is estimated to be $E_g = 0.134 \pm 0.034$ eV. This value is in good agreement with the band gap estimated from DFT calculations, which predict that a Dirac-like dispersion near the Fermi level is gapped by strong spin-orbit coupling of Ta and Te, resulting in a narrow gap of 0.10–0.15 eV[16]. Given that the valence-band maximum observed by ARPES is located at $E = E_F - 0.18$ eV, the conduction band minimum is expected to lie close to the Fermi level.

The low optimal operating temperature of Ta$_4$SiTe$_4$ is closely related to its narrow band gap. An empirical rule suggests that thermoelectric performance is often maximized at a temperature satisfying $E_g \approx 10 k_B T$. The maximum thermoelectric performance of Ta$_4$SiTe$_4$ is observed at around 200 K[16,23,24], which is consistent with this expectation when the experimentally observed band gap of approximately 0.134 eV is substituted into the relation. The Seebeck coefficient of Ta$_4$SiTe$_4$ reaches approximately −400 µV/K at 200 K, whereas its magnitude decreases to about −200 µV/K at 300 K[23]. As the temperature approaches the intrinsic-like regime near room temperature, thermally activated holes are expected to contribute to charge transport. The coexistence of electrons and holes likely reduces the absolute value of the Seebeck coefficient at 300 K through partial cancellation of their contributions.

Figure 4(d) shows the temperature dependence of the carrier mobilities parallel ($\mu_{//} = (en\rho_{//})^{-1}$) and perpendicular ($\mu_\perp = (en\rho_\perp)^{-1}$) to the $c$ axis, where $\rho_{//}$ and $\rho_\perp$ are taken from the data in Fig. 3(b). In the following, we discuss the carrier relaxation (scattering) mechanism based on the temperature dependence of the mobilities around 200 K. Figure 4(e) shows ln $\mu$ plotted as a function of ln $T$ in the temperature range from 140 to 240 K. The mobilities exhibit a power-law dependence on temperature, expressed as $\mu \propto T^p$, with fitting exponents $p_{//} = -1.68$ and $p_\perp = -1.54$ for directions parallel and perpendicular to the $c$ axis, respectively. The nearly identical temperature dependence of the mobilities indicates that the dominant scattering mechanisms are essentially isotropic in both directions. From these observations, we conclude that the one-dimensional anisotropy $\rho_\perp/\rho_{//} \approx 5.3$ around 200 K originates primarily from anisotropy in the band structure rather than from anisotropic scattering mechanisms. Notably, the relatively small resistivity anisotropy, on the order of unity, suggests a weakly one-dimensional transport character. This character is expected to be relatively robust against disorder, as supported by previous studies on solid solutions of Ta$_4$SiTe$_4$ and Nb$_4$SiTe$_4$[23,25]. In these systems, the transport properties are only weakly modified by chemical substitution, indicating a high tolerance of the conduction channels to disorder.

At 200 K (100 K), where Ta$_4$SiTe$_4$ exhibits giant thermoelectric performance, the carrier density is $n \approx 2.3 \times 10^{18}$ cm$^{-3}$ ($\approx 9 \times 10^{17}$ cm$^{-3}$), and the mobility parallel to the $c$ axis reaches $\mu_{//} \approx 600$ cm$^2 \cdot$V$^{-1} \cdot$s$^{-1}$ ($\approx 1700$ cm$^2 \cdot$V$^{-1} \cdot$s$^{-1}$). For comparison, in the typical room-temperature thermoelectric material Bi$_2$Te$_3$, the carrier density and mobility at 300 K are $n \approx 1.5 \times 10^{19}$ cm$^{-3}$ and $\mu \approx 200$ cm$^2 \cdot$V$^{-1} \cdot$s$^{-1}$, respectively[26]. This comparison indicates that the low resistivity of Ta$_4$SiTe$_4$ originates from the high mobility rather than from a high carrier density. This characteristic likely reflects anisotropic orbital bonding involving Ta and Te atoms along the $c$ axis, which gives rise to weakly one-dimensional conduction channels. As a result, the trade-off between the Seebeck coefficient and electrical conductivity is partially mitigated, leading to a high thermoelectric power factor.

In this study, we investigated the anisotropic electronic states underlying the high thermoelectric performance of the van der Waals material Ta$_4$SiTe$_4$. By combining micro-ARPES and FIB-based transport measurements, we revealed a one-dimensional electronic structure and a narrow band gap. This electronic nature enables the simultaneous realization of a large Seebeck coefficient and high electrical conductivity, while maintaining robust charge transport against disorder. As a result, Ta$_4$SiTe$_4$ exhibits an exceptionally large thermoelectric power factor at low temperatures. These findings clarify the correlation between weakly one-dimensional electronic structures and enhanced thermoelectric performance, and establish clear guidelines for the design and identification of high-performance thermoelectric materials.

**Author Contributions:** M.Has., Y.Ok., and H.I. conceived the experiment. F.M., Y.Ok., and K.T. grew the Ta$_4$SiTe$_4$ single crystals. M.Ho. performed the ARPES measurements with H.I, A.T., and F.K. H.I., M.K., M.I., and A.T. prepared the FIB samples. H.I. and A.T. conducted the electrical transport measurements. H.I., A.T., M.Har., Y.Y., and M.Ho. analyzed the data. H.I. and M.Has. wrote and revised the manuscript with input from A.T., M.Har., Y.Y., Y.Ok., M.Ho., and F.K. M.Has. supervised the project. All authors discussed the results and approved the final manuscript.

**Funding Sources:** This work was supported by Grants-in-Aid for Scientific Research (Grants No. JP22H00112, JP23H04866, JP23H04868, JP23K04579, JP24H00827, and JP25H00613), the Steel Foundation for Environmental Protection Technology, the CASIO SCIENCE PROMOTION FOUNDATION (J42-05), UTEC-UTokyo FSI Research Grant Program, Toray Science and Technology Grant, and the Japan Science and Technology Agency (JST) as part of Adopting Sustainable Partnerships for Innovative Research Ecosystem (ASPIRE), Grant Number JPMJAP2410 and JPMJAP2314. ARPES measurements were performed under the approval of the Photon Factory Program Advisory Committee (Proposal Nos. 2023G122 and 2025G050).

**Abbreviations:** FIB, focused ion beam; ARPES, angle-resolved photoemission spectroscopy

# REFERENCES

(1) MAHAN, G. D. Good Thermoelectrics. *Solid State Phys.* **1998**, *51*, 81–157. https://doi.org/10.1016/S0081-1947(08)60190-3.

(2) *Thermoelectrics Handbook*; Rowe, D. M., Ed.; CRC Press, 2018. https://doi.org/10.1201/9781420038903.

(3)  Terasaki, I. Transport Properties and Electronic States of the Thermoelectric Oxide NaCo$_2$O$_4$. *Phys. B Condens. Matter* **2003**, *328* (1–2), 63–67. https://doi.org/10.1016/S0921-4526(02)01810-0.

(4)  Takeuchi, T.; Kondo, T.; Takami, T.; Takahashi, H.; Ikuta, H.; Mizutani, U.; Soda, K.; Funahashi, R.; Shikano, M.; Mikami, M.; Tsuda, S.; Yokoya, T.; Shin, S.; Muro, T. Contribution of Electronic Structure to the Large Thermoelectric Power in Layered Cobalt Oxides. *Phys. Rev. B* **2004**, *69* (12), 125410. https://doi.org/10.1103/PhysRevB.69.125410.

(5)  Ohta, H.; Mune, Y.; Koumoto, K.; Mizoguchi, T.; Ikuhara, Y. Critical Thickness for Giant Thermoelectric Seebeck Coefficient of 2DEG Confined in SrTiO$_3$/SrTi$_{0.8}$Nb$_{0.2}$O$_3$ Superlattices. *Thin Solid Films* **2008**, *516* (17), 5916–5920. https://doi.org/10.1016/j.tsf.2007.10.034.

(6)  Zhao, L.-D.; Lo, S.-H.; Zhang, Y.; Sun, H.; Tan, G.; Uher, C.; Wolverton, C.; Dravid, V. P.; Kanatzidis, M. G. Ultralow Thermal Conductivity and High Thermoelectric Figure of Merit in SnSe Crystals. *Nature* **2014**, *508* (7496), 373–377. https://doi.org/10.1038/nature13184.

(7)  Huang, B.; Lawrence, C.; Gross, A.; Hwang, G.-S.; Ghafouri, N.; Lee, S.-W.; Kim, H.; Li, C.-P.; Uher, C.; Najafi, K.; Kaviany, M. Low-Temperature Characterization and Micropatterning of Coevaporated Bi$_2$Te$_3$ and Sb$_2$Te$_3$ Films. *J. Appl. Phys.* **2008**, *104* (11). https://doi.org/10.1063/1.3033381.

(8)  Li, H.; Feng, J.; Zhao, L.; Min, E.; Zhang, H.; Li, A.; Li, J.; Liu, R. Hierarchical Low-Temperature n-Type Bi$_2$Te$_3$ with High Thermoelectric Performances. *ACS Appl. Mater. Interfaces* **2024**, *16* (17), 22147–22154. https://doi.org/10.1021/acsami.4c02141.

(9) Chung, D.-Y.; Hogan, T.; Brazis, P.; Rocci-Lane, M.; Kannewurf, C.; Bastea, M.; Uher, C.; Kanatzidis, M. G. $CsBi_4Te_6$: A High-Performance Thermoelectric Material for Low-Temperature Applications. *Science (80-. ).* **2000**, *287* (5455), 1024–1027. https://doi.org/10.1126/science.287.5455.1024.

(10) Chung, D.-Y.; Hogan, T. P.; Rocci-Lane, M.; Brazis, P.; Ireland, J. R.; Kannewurf, C. R.; Bastea, M.; Uher, C.; Kanatzidis, M. G. A New Thermoelectric Material: $CsBi_4Te_6$. *J. Am. Chem. Soc.* **2004**, *126* (20), 6414–6428. https://doi.org/10.1021/ja039885f.

(11) Nakano, A.; Maruoka, U.; Kato, F.; Taniguchi, H.; Terasaki, I. Room Temperature Thermoelectric Properties of Isostructural Selenides $Ta_2PdS_6$ and Ta2PdSe6. *J. Phys. Soc. Japan* **2021**, *90* (3), 033702. https://doi.org/10.7566/JPSJ.90.033702.

(12) Nakano, A.; Yamakage, A.; Maruoka, U.; Taniguchi, H.; Yasui, Y.; Terasaki, I. Giant Peltier Conductivity in an Uncompensated Semimetal $Ta_2PdSe_6$. *J. Phys. Energy* **2021**, *3* (4), 044004. https://doi.org/10.1088/2515-7655/ac2357.

(13) Nakano, A.; Maruoka, U.; Terasaki, I. Correlation between Thermopower and Carrier Mobility in the Thermoelectric Semimetal $Ta_2PdSe_6$. *Appl. Phys. Lett.* **2022**, *121* (15). https://doi.org/10.1063/5.0102434.

(14) Badding, M. E.; DiSalvo, F. J. Synthesis and Structure of $Ta_4SiTe_4$, a New Low-Dimensional Material. *Inorg. Chem.* **1990**, *29* (20), 3952–3954. https://doi.org/10.1021/ic00345a009.

(15) Li, J.; Hoffmann, R.; Badding, M. E.; DiSalvo, F. J. Electronic and Structural Properties of the Novel Chain Compound Tantalum Telluride Silicide, $Ta_4Te_4Si$. *Inorg. Chem.* **1990**, *29* (20), 3943–3952. https://doi.org/10.1021/ic00345a008.

(16) Inohara, T.; Okamoto, Y.; Yamakawa, Y.; Yamakage, A.; Takenaka, K. Large Thermoelectric Power Factor at Low Temperatures in One-Dimensional Telluride Ta$_4$SiTe$_4$. *Appl. Phys. Lett.* **2017**, *110* (18). https://doi.org/10.1063/1.4982623.

(17) Okamoto, Y.; Yoshikawa, Y.; Wada, T.; Takenaka, K. Hole-Doped M$_4$SiTe$_4$ ( M = Ta, Nb) as an Efficient p -Type Thermoelectric Material for Low-Temperature Applications. *Appl. Phys. Lett.* **2019**, *115* (4). https://doi.org/10.1063/1.5109590.

(18) Xu, Q.; Qu, S.; Ming, C.; Qiu, P.; Yao, Q.; Zhu, C.; Wei, T.-R.; He, J.; Shi, X.; Chen, L. Conformal Organic–Inorganic Semiconductor Composites for Flexible Thermoelectrics. *Energy Environ. Sci.* **2020**, *13* (2), 511–518. https://doi.org/10.1039/C9EE03776D.

(19) Liu, M.; Ren, D.; Ye, C.; Yin, T.; Qu, S.; Zong, P. Enhanced Thermoelectric Performance of (Ta$_{1-x}$Mo$_x$)$_4$SiTe$_4$/Polyvinylidene Fluoride (PVDF) Organic–Inorganic Flexible Thermoelectric Composite Films. *Nanoscale* **2025**, *17* (19), 12441–12449. https://doi.org/10.1039/D5NR00816F.

(20) Liu, S.; Yin, H.; Singh, D. J.; Liu, P.-F. Ta$_4$SiTe$_4$ : A Possible One-Dimensional Topological Insulator. *Phys. Rev. B* **2022**, *105* (19), 195419. https://doi.org/10.1103/PhysRevB.105.195419.

(21) Matsunaga, F.; Okamoto, Y.; Yokoyama, Y.; Takehana, K.; Imanaka, Y.; Nakamura, Y.; Kishida, H.; Kawano, S.; Matsuhira, K.; Takenaka, K. Anisotropic Optical Conductivity Accompanied by a Small Energy Gap in the One-Dimensional Thermoelectric Telluride Ta$_4$SiTe$_4$. *Phys. Rev. B* **2024**, *109* (16), L161105. https://doi.org/10.1103/PhysRevB.109.L161105.

(22) Kitamura, M.; Souma, S.; Honma, A.; Wakabayashi, D.; Tanaka, H.; Toyoshima, A.; Amemiya, K.; Kawakami, T.; Sugawara, K.; Nakayama, K.; Yoshimatsu, K.;

Kumigashira, H.; Sato, T.; Horiba, K. Development of a Versatile Micro-Focused Angle-Resolved Photoemission Spectroscopy System with Kirkpatrick–Baez Mirror Optics. *Rev. Sci. Instrum.* **2022**, *93* (3). https://doi.org/10.1063/5.0074393.

(23) Okamoto, Y.; Wada, T.; Yamakawa, Y.; Inohara, T.; Takenaka, K. Large Thermoelectric Power Factor in One-Dimensional Telluride $Nb_4SiTe_4$ and Substituted Compounds. *Appl. Phys. Lett.* **2018**, *112* (17). https://doi.org/10.1063/1.5023427.

(24) Tomitaka, G.; Kawano, S.; Okamoto, Y.; Matsuhira, K. Optimal Carrier Concentration of Low-Temperature Thermoelectric Material $M_4SiTe_4$ ( M = Ta, Nb) from First-Principles Calculations. *J. Phys. Soc. Japan* **2025**, *94* (11). https://doi.org/10.7566/JPSJ.94.114705.

(25) Yoshikawa, Y.; Wada, T.; Okamoto, Y.; Abe, Y.; Takenaka, K. Large Thermoelectric Power Factor in Whisker Crystals of Solid Solutions of the One-Dimensional Tellurides $Ta_4SiTe_4$ and $Nb_4SiTe_4$. *Appl. Phys. Express* **2020**, *13* (12), 125505. https://doi.org/10.35848/1882-0786/abcc3d.

(26) Akhanda, M. S.; Rezaei, S. E.; Esfarjani, K.; Krylyuk, S.; Davydov, A. V.; Zebarjadi, M. Thermomagnetic Properties of $Bi_2Te_3$ Single Crystal in the Temperature Range from 55 K to 380 K. *Phys. Rev. Mater.* **2021**, *5* (1), 015403. https://doi.org/10.1103/PhysRevMaterials.5.015403.

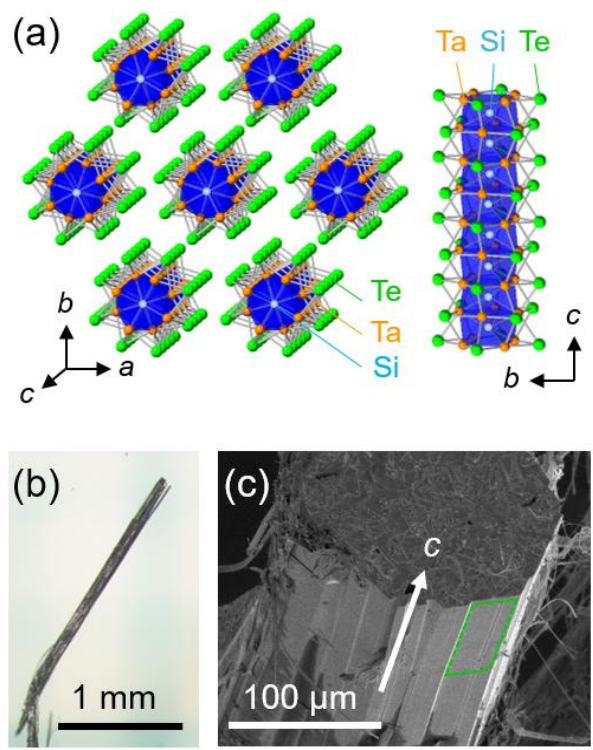

**Figure 1:** Crystal structure of Ta$_4$SiTe$_4$. (a) Schematic three-dimensional representation of the crystal structure of Ta$_4$SiTe$_4$, in which the one-dimensional chains run parallel to the *c* axis. (b) Optical microscope image of a Ta$_4$SiTe$_4$ whisker crystal. (c) Scanning electron microscope picture of a Ta$_4$SiTe$_4$ crystal. The *c* axis is indicated by a white arrow. The region enclosed by the dotted green line was picked up by FIB fabrication for the measurements represented in Figs. 3 and 4.

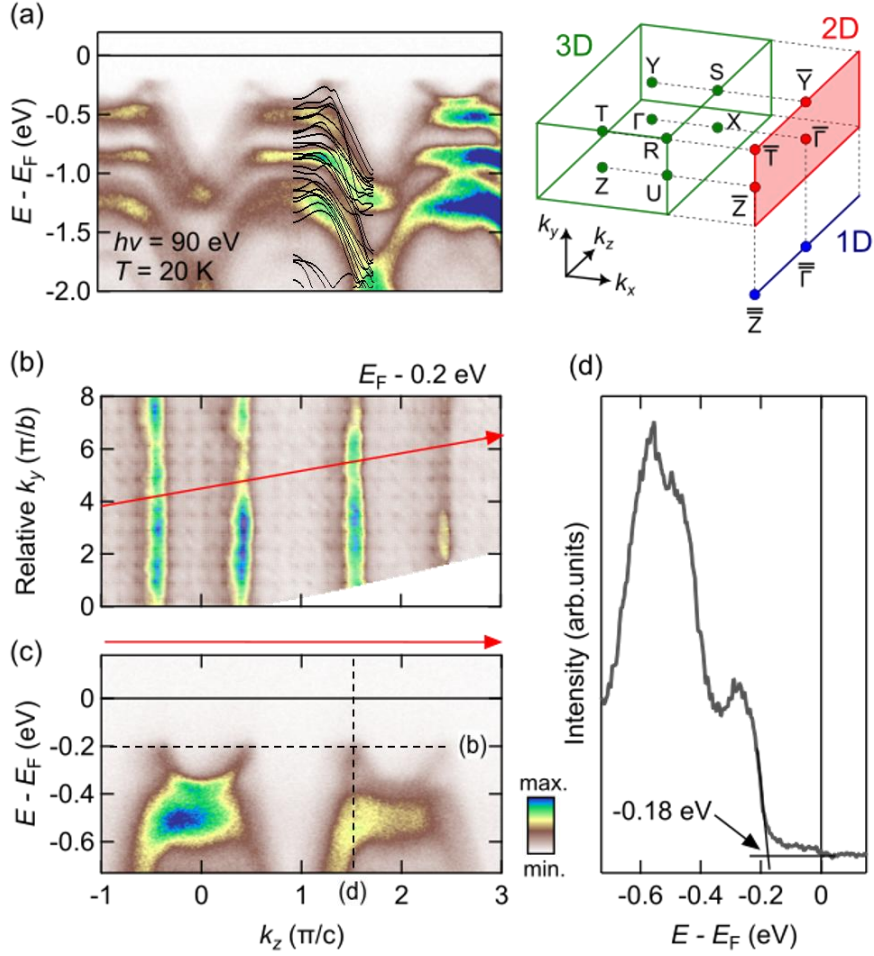

**Figure 2.** ARPES measurements of Ta$_4$SiTe$_4$. (a) ARPES spectrum along the $\bar{\bar{\Gamma}}$–$\bar{\bar{Z}}$ direction measured at $hv$ = 90 eV and 20 K, where $\bar{\bar{\Gamma}}$ and $\bar{\bar{Z}}$ denote the high-symmetry points $\Gamma$ and Z projected onto the $k_z$ direction (parallel to the $c$ axis). A schematic of the Brillouin zone is shown on the right. The black lines represent band dispersion calculated in the previous work[16]. (b) Constant-energy map in the $k_y$–$k_z$ plane at $E = E_F - 0.2$ eV, where $k_y$ is shown as a relative value. (c) ARPES spectrum along the red line in panel (b). (d) The photoemission intensity as a function of energy along the vertical dashed line in (c).

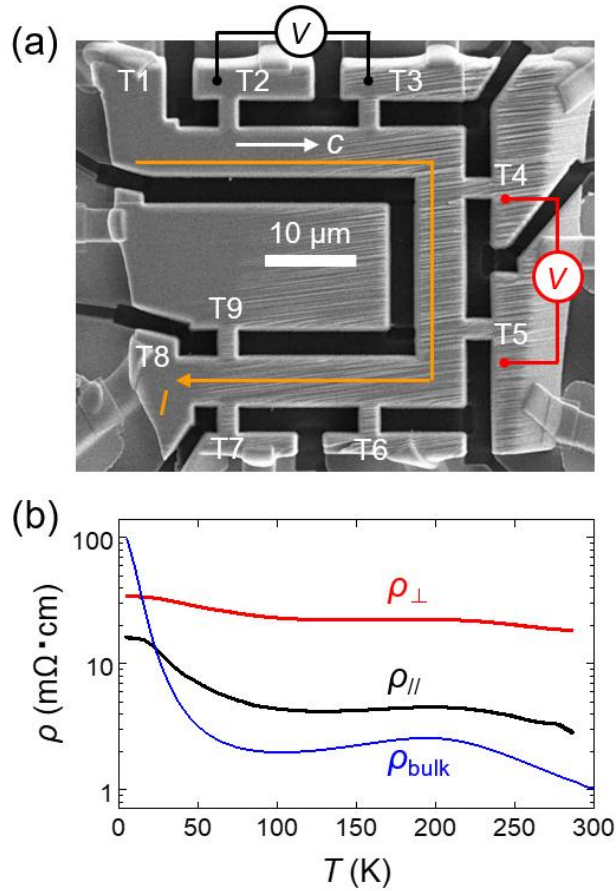

**Figure 3:** Anisotropy of the electrical resistivity in $Ta_4SiTe_4$. (a) Scanning electron microscope image of an FIB-fabricated $Ta_4SiTe_4$ sample. Configurations of the electrical measurement are drawn. T1–T9 denote the terminal numbers used for the electrical contacts. (b) Resistivity parallel (black) and perpendicular (red) to the $c$ axis as a function of the temperature. Blue line indicates the result for a bulk sample.

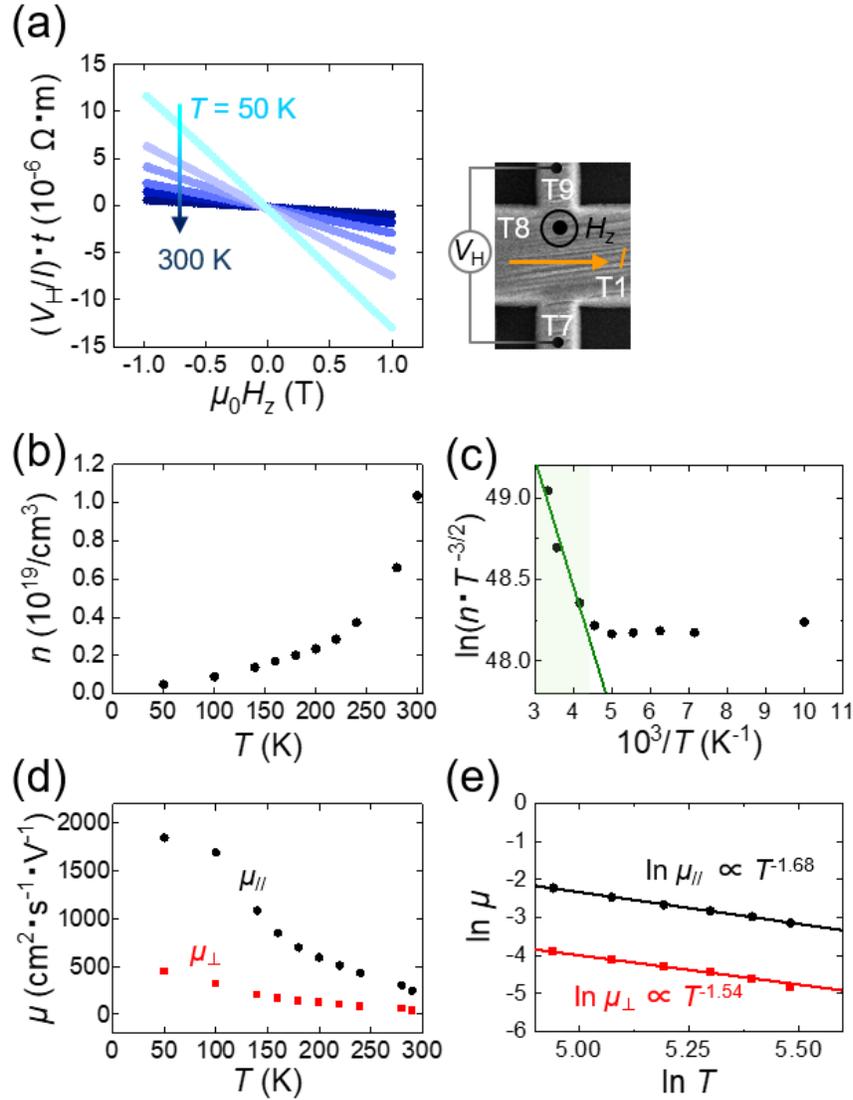

**Figure 4:** Hall effect of $Ta_4SiTe_4$. (a) Results of Hall measurements at various temperatures. (b) Temperature dependence of the carrier density determined by Hall measurements. (c) Plot of ln ($n \cdot T^{-3/2}$) as a function of $T^{-1}$. Green fill represents assumed intrinsic-like region. Green line is the fit by Eq. (2). (d) Temperature dependence of the mobility parallel (black) and perpendicular (red) to the $c$ axis. (e) Plot of ln $\mu$ as a function of ln$T$ in the range of $T = 140 – 240$ K. Lines are the fit by a straight line.